\documentclass{article}
\usepackage{spconfa4,amsmath,graphicx}

\usepackage{cite}
\usepackage{amsmath,amssymb,amsfonts}
\usepackage{algorithmic}
\usepackage{graphicx}
\usepackage{textcomp}
\usepackage{glossaries}
\usepackage{xcolor}
\usepackage{graphicx}
\usepackage{psfrag}
\usepackage{siunitx}
\usepackage{bm,bbm}
\usepackage{mathtools}
\usepackage{booktabs}
\usepackage[capitalise]{cleveref}
\usepackage{url}
\usepackage{multirow}
\usepackage{makecell}
\usepackage{pifont}
\usepackage{adjustbox}

\usepackage{amssymb}% http://ctan.org/pkg/amssymb
\usepackage{pifont}% http://ctan.org/pkg/pifont

\usepackage{tikz,pgfplots}
\usepackage{tikzscale}

\pgfplotsset{compat=1.11,
    /pgfplots/ybar legend/.style={
    /pgfplots/legend image code/.code={%
       \draw[##1,/tikz/.cd,yshift=-0.25em]
        (0cm,0cm) rectangle (3pt,0.8em);},
   },
}

\usepgfplotslibrary{groupplots}

\usepackage{microtype}
\usepackage{cleveref}
\usepackage{multirow}

\newacronym{DNN}{DNN}{deep neural network}
\newacronym{CDR}{CDR}{coherent-to-diffuse power ratio}
\newacronym{DRR}{DRR}{direct-to-reverberant energy ratio}
\newacronym{DoA}{DoA}{direction-of-arrival}
\newacronym{TDoA}{TDoA}{time difference of arrival}
\newacronym{MSE}{MSE}{mean-squared error}
\newacronym{MLP}{MLP}{multilayer perceptron}
\newacronym{CRNN}{CRNN}{convolutional recurrent neural network}
\newacronym[plural=GPs,firstplural=Gaussian processes (GPs)]{GP}{GP}{Gaussian process}
\newacronym{GRU}{GRU}{gated recurrent unit}
\newacronym{RIR}{RIR}{room impulse response}
\newacronym{AWGN}{AWGN}{additive white Gaussian noise}
\newacronym{SNR}{SNR}{signal-to-noise ratio}
\newacronym{STFT}{STFT}{short-time Fourier transform}
\newacronym[plural=PSDs,firstplural=power spectral densities~(PSDs)]{PSD}{PSD}{power spectral density}
\newacronym{AE}{AE}{absolute error}
\newacronym{MAE}{MAE}{mean-absolute error}
\newacronym{PE}{PE}{position error}
\newacronym{MPE}{MPE}{mean position error}
\newacronym{WASN}{WASN}{wireless acoustic sensor network}
\newacronym{OoR}{OoR}{out-of-range}
\newacronym{ASR}{ASR}{automatic speech recognition}
\newacronym{TDNN}{TDNN}{time delay neural network}
\newacronym{CNN}{CNN}{convolutional neural network}
\newacronym{WLS}{WLS}{weighted least squares}
\newacronym{LS}{LS}{least squares}
\newacronym{RANSAC}{RANSAC}{random sample consensus}
\newacronym{MVDR}{MVDR}{minimum variance distortionless response}

\newacronym{GARDE}{GARDE}{\textbf{G}eometry c\textbf{A}libration f\textbf{R}om \textbf{D}istance \textbf{E}stimates}
\newacronym{MDS}{MDS}{Multi Dimensional Scaling}

\newacronym{CWLS}{CWLS}{constrained weighted least squares}
\newacronym{CRLB}{CRLB}{Cramer-Rao lower bound}
\newacronym{RMSE}{RMSE}{root mean square error}

\newacronym{CDF}{CDF}{cumulative distribution function}
\newacronym{CDF2}{CDF}{Cumulative distribution function}

\newacronym{ABEL}{ABEL}{averaged burst error length}
\newacronym{ACD}{ACD}{average coherence drift}
\newacronym{ADC}{ADC}{analog-digital converter}
\newacronym{APT}{APT}{averaged processing time}
\newacronym{ASN}{ASN}{acoustic sensor network}
\newacronym{ASNs}{ASNs}{acoustic sensor networks}
\newacronym{ASRC}{ASRC}{arbitrary sampling rate conversion}
\newacronym{ATD}{ATD}{time drift}
\newacronym{ATS}{ATS}{accumulating time shift}
\newacronym{BI}{BI}{band-limited interpolation}
\newacronym{BSS}{BSS}{blind source separation}
\newacronym{CCF}{CCF}{cross-correlation function}
\newacronym{CCF-2}{CCF-2}{secondary \gls{CCF}}
\newacronym{CD}{CD}{coherence drift}
\newacronym{CFM}{CFM}{coherence function maximization}
\newacronym{CM}{CM}{correlation maximization}
\newacronym{CSD}{CSD}{cross-spectral density}
\newacronym{CSD-2}{CSD-2}{secondary \gls{CSD}}
\newacronym{CTC}{CTC}{continuous-time conversion}
\newacronym{DCML}{DCML}{\gls{ML} method for dynamic conditions}
\newacronym{DB}{DB}{Data base}
\newacronym{DD}{DD}{digital-to-digital}
\newacronym{DDC}{DDC}{digital-to-digital converter}
\newacronym{DSP}{DSP}{digital signal processing}
\newacronym{DTC}{DTC}{discrete-time conversion}
\newacronym{DXCP}{DXCP}{double-cross-correlation processor}
\newacronym{DXCPPhaT}{DXCP-PhaT}{\gls{DXCP} with phase transform}
\newacronym{ECS}{ECS}{energy correlation score}
\newacronym{FCF}{FCF}{filter correlation function}
\newacronym{FF}{FF}{free-field}
\newacronym{FFT}{FFT}{fast Fourier transform}
\newacronym{FFT-DXCP}{FFT-DXCP}{\gls{FFT} domain \gls{DXCP}}
\newacronym{FO}{FO}{frame-oriented}
\newacronym{GCC}{GCC}{generalized cross-correlation}
\newacronym{GCPSD}{GCPSD}{generalized cross power spectral density}
\newacronym{GCCF}{GCCF}{generalized \gls{CCF}}
\newacronym{GCCPhaT}{GCC-PhaT}{generalized cross-correlation with phase transform}
\newacronym{GCSD}{GCSD}{generalized cross-spectral density}
\newacronym{GE}{GE}{Gilbert-Elliott}
\newacronym{IBI}{IBI}{iterative \gls{BI}}
\newacronym{ICA}{ICA}{independent component analysis}
\newacronym{IML}{IML}{iterative \gls{ML}}
\newacronym{IR}{IR}{impulse response}
\newacronym{IIR}{IIR}{infinite impulse response}
\newacronym{IFFT}{IFFT}{inverse fast Fourier transform}
\newacronym{ISTFT}{ISTFT}{inverse \gls{STFT}}
\newacronym{LCD}{LCD}{least-squares coherence drift}
\newacronym{LPD}{LPD}{linear-phase drift}
\newacronym{LTI}{LTI}{linear time-invariant}
\newacronym{LTV}{LTV}{linear time-variant}
\newacronym{LTVIR}{LTV-IR}{linear time-variant impulse response}
\newacronym{ML}{ML}{maximum likelihood}
\newacronym{MS}{MS}{multi-stage}
\newacronym{MSC}{MSC}{magnitude-squared coherence}
\newacronym{OML}{OML}{optimized \gls{ML}}
\newacronym{OR}{OR}{outlier removal}
\newacronym{PHAT}{PHAT}{phase transform}
\newacronym{PL}{PL}{packet loss}
\newacronym{PolyFar}{PolyFar}{polyphase-Farrow}
\newacronym{ppb}{ppb}{part per billion}
\newacronym{ppm}{ppm}{parts \ per \ million}

\newacronym{RBI}{RBI}{recursive band-limited interpolation}
\newacronym{RF}{RF}{realtime factor}
\newacronym{RTP}{RTP}{Real-time Transport Protocol}
\newacronym{RTCP}{RTCP}{Real-time Transport Control Protocol}

\newacronym{SCOT}{SCOT}{smoothed coherence transform}
\newacronym{SINR}{SINR}{signal-to-interpolation-noise ratio}
\newacronym{SO}{SO}{sample-oriented}
\newacronym{SPIB}{SPIB}{signal processing information base}
\newacronym{SPL}{SPL}{signal packet loss}
\newacronym{SRC}{SRC}{sampling rate conversion}
\newacronym{SRO}{SRO}{sampling rate offset}
\newacronym{SCM}{SCM}{spatial covariance matrix}
\newacronym{STO}{STO}{sampling time offset}
\newacronym{TCP}{TCP}{Transmission Control Protocol}
\newacronym{TD}{TD}{time domain}
\newacronym{TDDXCP}{TD-DXCP}{time domain \gls{DXCP}}
\newacronym{TDD}{TDD}{time-delay difference}
\newacronym{TDOA}{TDOA}{time-difference of arrival}
\newacronym{UDP}{UDP}{User Datagram Protocol}
\newacronym{VAD}{VAD}{activity detection}
\newacronym{SAD}{SAD}{sound activity detection}
\newacronym{WACD}{WACD}{weighted average coherence drift}
\newacronym{DWACD}{DWACD}{dynamic weighted average coherence drift}
\newacronym{WG}{WG}{weighting}
\newacronym{WLCD}{WLCD}{weighted \gls{LCD}}
\newacronym{WASNs}{WASNs}{wireless acoustic sensor networks}
\newacronym{WCP}{WCP}{wideband correlation processor}
\newacronym{XC}{XC}{cross-correlation}
\newacronym{SVD}{SVD}{singular value decomposition}

\newacronym{TXCO}{TXCO}{temperature compensated crystal oscillator}

\newacronym{OU}{OU}{Ornstein-Uhlenbeck}

\newacronym[firstplural=time differences of flight (TDOFs)]{TDOF}{TDOF}{time difference of flight}

\newacronym{cACGMM}{cACGMM}{complex Angular Central Gaussian Mixture Model}
\newacronym{EM}{EM}{Expectation Maximization}
\newacronym{EVD}{EVD}{eigenvalue decomposition}

\newacronym{cpWER}{cpWER}{concatenated minimum-permutation word error rate}
\newacronym{WER}{WER}{word error rate}

\newacronym{SDR}{SDR}{signal-to-distortion ratio}
\newacronym{ISM}{ISM}{image source method}
\newacronym{ILD}{ILD}{inter-level differences}
\newacronym{SWA}{SWA}{stochastic weight averaging}
\newacronym{PRA}{PRA}{pyroomacoustics}
\newcommand{\epa}[1]{\hat{\varepsilon}_{12}[\tilde{\ell'}]}

\usepackage{tikzscale}
\usepackage{tikz-layers}

\title{Diminishing Domain Mismatch for DNN-Based Acoustic Distance Estimation via Stochastic Room Reverberation Models}
\name{Tobias Gburrek, Adrian Meise, Joerg Schmalenstroeer and Reinhold Haeb-Umbach}
\address{Paderborn University, Department of Communications Engineering, Germany \\
  \{gburrek, schmalen, haeb\}@nt.uni-paderborn.de}

\begin{document}
\ninept

\maketitle
\begin{abstract}
The \gls{RIR} encodes, among others, information about the distance of an acoustic source from the sensors.  \Glspl{DNN} have been shown to be able to extract that information for acoustic distance estimation. Since there exists only a very limited amount of annotated data, e.g., \mbox{\glspl{RIR}} with distance information, training a DNN for acoustic distance estimation has to rely on simulated RIRs, resulting in an unavoidable mismatch to RIRs of real rooms. In this contribution, we show that this mismatch can be reduced by a novel combination of geometric and stochastic modeling of RIRs, resulting in a significantly improved distance estimation accuracy.
\end{abstract}
\glsresetall

%% Tex magic
% !TeX spellcheck = en_US
% !TeX encoding = utf-8
% !TeX root = ./../main.tex
% !TeX program = pdflatex 

\section{Introduction}
Knowing the distance between a speaker and the recording device can be valuable information for downstream signal processing tasks, e.g., for geometry calibration in \glspl{WASN} \cite{GeoJournal}, signal processing in hearing aids \cite{drr_estimate} or source extraction \cite{petermann2024}.
Common approaches estimate the distance between an acoustic source and a compact recording device with multiple microphones by evaluating the power ratio between the coherent signal part, originating from the direct signal propagation path, and the diffuse signal part which summarizes the propagation paths with multiple reflections  \cite{brendel_distributed_2019, brendel_probabilistic_2019}. 

In real environments each room has individual acoustic transfer functions, that depend not only on the distance between the recording device and the acoustic source but also on the room's shape, the positions of the device and the source, furniture, and materials on walls, ceiling and floor. Hence, either training data of the room under consideration or at least data from rooms with similar characteristics are required to finetune the parameters of a distance estimator and thus increase the model's precision \cite{Kushwaha23, Neri23, Neri24}.

Collecting and annotating recordings from real environments with diverse room sizes and reverberation conditions is a tedious task. Publicly available data is usually limited in one of the required variabilities: Meeting data often lacks ground truth positioning information, while data intended for comparing localization techniques usually stem only from a very limited number of rooms. 
As shown in~\cite{Kushwaha23}, this limited size of these data sets also limits the performance of data-driven distance estimation methods. 
%To our knowledge, there is no data set that provides a variety of recordings from multiple rooms with annotated distance information and a variety of reverberation times. \fromtgb{STARSS23 has 16 rooms}

Recent approaches to distance estimation are based on \glspl{DNN}, be it single-channel \cite{Neri24} or multi-channel \cite{distance_estimator}, \mbox{and require  a large amount of training data to} reliably generalize to unknown data. In \cite{distance_estimator} we proposed to use a \gls{CRNN} trained on simulated data, which leveraged the problem by using diverse room setups for generating training data, that were similar to the rooms during tests. To this end, synthetic room impulse responses were generated via an \gls{ISM} \cite{HP_Allen_Berkley} and subsequently convolved with speech data. This synthetic data models sources and microphones with omnidirectional characteristics, which directly influences the distance-related features, e.g., \gls{CDR}, and thus leads to a mismatch between synthetic data and recordings from real environments. A \gls{DNN} trained with synthetic \glspl{RIR} with omnidirectional sources will have difficulties dealing with real-world data with typically directional sources~\cite{GeoJournal, Neri24}.
To reduce the resulting systematic errors on real-world data, \gls{DRR} augmentation techniques can be used~\cite{Bryan20, GeoJournal}. 
Alternatively, synthetic data may be enriched by recordings from real environments or pre-trained models may be fine-tuned to environments, as for example proposed by the authors of \cite{Kushwaha23}. 
However, the generalization ability between different data sets typically is limited as shown in~\cite{Neri24}. 
%(LOCATA) and \cite{politis2022starss22} (STARSS) \fromtgb{The authors of LOCATA and STARSS did not propose this! cite dist es paper.}. 
%Although this approach can improve performance on real-world data, it requires recording with hardware that has a similar microphone spacing to the microphone array in the application, which is obviously in limited availability. \fromtgb{Single channel dist est? More general: This a solution for a single data set. This is like the room specific trainig stage of the Gaussian Processes from FAU.}

In order to be able to create large synthetic data sets for training a distance estimator, the modeling of the \glspl{RIR} must become more realistic and, for example, include directional characteristics of sources and microphones to enable the applicability of the models to arbitrary scenarios. For example, there are approaches such as \cite{TS-RIR} that directly learn to map synthetic \glspl{RIR} to real \glspl{RIR}. However, the approach from  \cite{TS-RIR} is not suitable for the problem at hand, since the general structure of a real \gls{RIR} may be adapted, but the exact parameters of the simulated scenario, e.g., the distance, are not preserved.

In this paper, we propose an approach to \gls{RIR} simulation with the aim of improving the performance of a data-driven distance estimator, trained with synthetic \glspl{RIR}, on data from real environments. 
While the signal propagation paths with only a few reflections are simulated using a geometric approach to model the mainly  specular characteristics of the reflection, the signal propagation paths with more reflections, which are mainly diffuse, are simulated based on a stochastic approach. 
Thereby, the power of the stochastic part of the simulated \glspl{RIR} is chosen so that the resulting \gls{DRR}, as distance carrying information, meets the relation between the source-microphone distance and the critical distance, which results from the parameters of the simulated room.
Furthermore, the directivity of the sources is taken into account in the geometrical part of the simulated \gls{RIR} and the calculation of the power of the stochastic part. 
%Hereby, we try to narrow the gap between real \glspl{RIR} and simulated ones by combining geometrical and stochastic simulation methods for \glspl{RIR} in a computationally efficient simulation strategy.  
Experiments have shown that training a \gls{DNN}-based distance estimator solely on the proposed simulated data improves its generalization ability to real data from the MIRaGe \cite{MIRaGe} and MIRD \cite{mird} data sets.

The paper is organized as follows: In Sec.~\ref{Sec:SimData} we briefly review  common techniques for simulating \glspl{RIR} before we present our approach to  generating \glspl{RIR} in Sec.~\ref{Sec:OurRIR}. After a short explanation of the used distance estimator in Sec.~\ref{Sec:DistEst}, experimental results are presented and discussed in Sec.~\ref{Sec:Experiments}. Finally, we end with some conclusions in Sec.~\ref{Sec:Conclusions}.
%% Tex magic
% !TeX spellcheck = en_US
% !TeX encoding = utf-8
% !TeX root = ./../main.tex
% !TeX program = pdflatex 

\section{Review on RIR simulation techniques}  \label{Sec:SimData}
Common simulation software for RIRs employs either the image source method, that approximately considers the geometrical setup of microphones and sound sources in a shoe box-shaped room \cite{habets_rir,pra, Xu24}, ray/cone tracing algorithms  utilizing 3D models \cite{Glitza23} or a combination of both. Although ray/cone tracing algorithms promise more realistic simulations than the image source method by considering furniture and different wall materials, it remains a tool for special purposes. The generation of diverse and detailed 3D models is time-consuming and the computational complexity of calculating the reflections and tracing the sound geometrically is intractable for the large amount of data required for \gls{DNN} training.

\subsection{Directivity of sources and microphones}
%In real-world applications sources of acoustic events, e.g., human speakers or devices, emit an acoustic signal at an unknown position in the room, and the recording device, e.g., a \gls{WASN} node, equipped with multiple microphones records this signal to estimate the distance to the source depending on the reverberation characteristics. 
Many acoustic sources   have a directivity pattern that significantly differs from an omnidirectional directivity pattern. As reported in \cite{Poerschmann2020} the directivity pattern of human speakers is frequency-dependent and depends on the type (vocal or fricative) of uttered phonemes. It can be roughly approximated by a cardioid characteristic, which can also be found in monitor loudspeakers.

This implies that depending on the direction of view the acoustic source and each mirrored image of the source get an extra image-dependent weighting factor in the image source method \cite{Hafezi15}. So the summation of all weighted image signals impinging on the microphone's position is taken as an approximate recording of a directive audio source. If the microphones also have a directivity pattern the impinging mirror signals have to be weighted in accordance to the direction of view of the microphone.

Although the image method has proven its usefulness in many publications, it tends to deliver sparse sequences of impulses that, when convolved with clean audio snippets, do not provide a natural sound perception for the human listener. A \gls{RIR} recorded in a real environment shows a much noisier and random structure than the \glspl{RIR} generated by the image method, especially for the late reverberation.

\subsection{Stochastic RIR models}
\label{subsec:stochasitc}
Some approaches, e.g., \cite{stochastic_model}, suggest to model \glspl{RIR} statistically as a random process with an exponentially decaying envelope, that is influenced by some basic acoustic parameters, to better capture general characteristics of a \gls{RIR} and to ignore the exact geometrical propagation of the sound. We extend the model from \cite{stochastic_model} with a delay $N_d$, i.e., the integer rounded time of flight between the acoustic source and the microphone, and approximate the RIR $h_s(n)$ by an exponentially decaying Gaussian process:
\begin{align}
  h_s(n) := \begin{cases}
  b(n) \: e^{-\Delta \frac{n-N_d}{f_s}} & \text{for } n \geq N_d \\
  0  & \text{else}
  \end{cases}  
  \label{stoch_h}
\end{align}
with $n$ as the sample index, $\Delta{=}3 {\cdot} \mathrm{ln}\left(10\right){/}T_{60}$, $f_s$ as the sampling frequency and $b(n)$ as zero-mean, white Gaussian noise with variance $\sigma_b^2$, i.e., $b(n) \sim \mathcal{N}(0,\sigma_b^2)$. 
However, if the model should reflect distance information and simultaneously consider the directivity patterns of the source and microphone, it has to be further extended.

\section{Proposed RIR simulation technique} \label{Sec:OurRIR}
As mentioned in~\cite{Schrder:50580} the image source method is suitable to model the early reflections of sound, which are mainly specular, but the late reflections, which are mainly diffuse, are not modeled appropriately by the image source method.
Hence, we propose to model only the early reflections using the image source method while the late reflections are modeled using the stochastic approach presented in~\cref{subsec:stochasitc}.
\begin{figure}[tb]
	\centering
	\resizebox{\linewidth}{!}{
        \Large
		\input{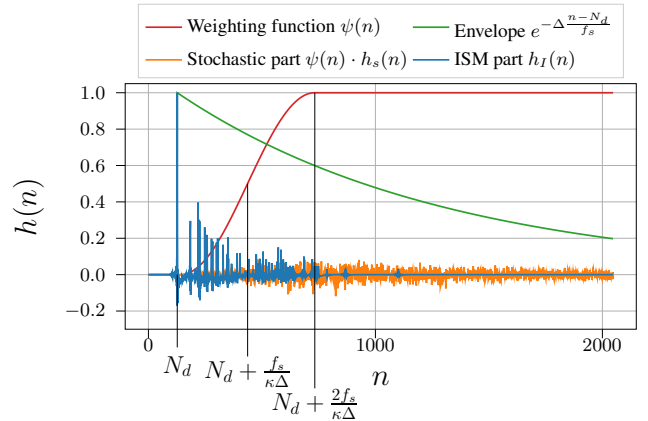}
    }
   	\caption{Visualization of the proposed approach to \gls{RIR} simulation}	
    \label{fig:simulation_approach}
\end{figure}
Therefore, we combine the image source method and the stochastic model of \eqref{stoch_h} to generate a \gls{RIR} $h(n)$ of length $N$ as follows (see Fig.~\ref{fig:simulation_approach})\footnote{Code is available at https://github.com/fgnt/paderwasn}: First, we simulate the early part of the \gls{RIR} $h_I(n)$ based on the image source method using the image sources up to order $K$. Thereby, a cardioid pattern is utilized for the source's directivity. 
We choose $K{=}3$ since reflections of higher order are nearly completely diffuse as reported in \cite{Schrder:50580}.
Additionally, a high pass filter is applied to the early part of the \gls{RIR} $h_I(n)$ as proposed in \cite{HP_Allen_Berkley}.

The diffuse, late reflections should follow \eqref{stoch_h} while preserving the distinct reflections modeled by $h_I(n)$. This is achieved by weighting $h_{s}(n)$ with the function
\begin{align}
  \psi(n) {=} \begin{cases}
  0 & \phantom{\text{for }}\:\:\: n {\leq} N_d \\
  \frac{1}{2}\mathopen{}\left(1 {-} \cos\mathopen{}\left(\frac{\pi (n-N_d)}{2 f_s/(\kappa \Delta)}\right)\mathclose{}\right)\mathclose{} & \text{for }\:\:\: N_d{<}n {\leq}  N_d {+} \frac{2 f_s}{\kappa \Delta}. \\
  %\sin^2\left(\frac{\pi (n-N_d)}{ f_s/(\kappa \Delta)}\right) & \text{for }\:\:\: N_d {<} n {\leq}  N_d {+} \frac{2 f_s}{\kappa \Delta}. \\
  1 & \phantom{\text{for }}\:\:\: n {>} N_d {+} \frac{2 f_s}{\kappa \Delta} 
  \end{cases}  
\end{align}
First experiments on the choice of $\kappa$ have shown that a better generalization to recorded \glspl{RIR} is achieved for $\kappa{=}1$.
This choice results in a smoother fade-in of the stochastic part of the \gls{RIR}, i.e., less disturbance of the early reflections.
The resulting \gls{RIR} $h(n)$ is given by
\begin{align}
    h(n) = h_I(n) + \psi(n) \cdot h_s(n).
\end{align}

Finally, the power of the Gaussian process in~\eqref{stoch_h} is chosen so that the \gls{RIR} $h(n)$ exhibits a desired \gls{DRR}. 
The desired \gls{DRR} $\eta$ is calculated based on the relation between geometrical as well as acoustic properties of the room  and the distance between the source and the microphone~\cite{drr_estimate}.
In addition to that, we extend the relation from~\cite{drr_estimate} by taking into account that the direct path component of the \gls{RIR} is scaled by the source's  directional response $D(\varphi, \varrho)$, where $\varphi$ and $\varrho$ are the azimuth and elevation angles between look direction of the source and relative position of the microphone, respectively. 
Thus, the desired \gls{DRR}  is given by
\begin{align}
  \eta = D^2(\varphi, \varrho) \cdot \frac{d^2_c}{d^2}
  \label{hat_d},
\end{align}
with $d$ as the distance between the source and microphone. 
$d_c$ denotes the critical distance with
\begin{align}
    d_c = \SI{0.1}{m} \cdot \sqrt{\alpha \cdot \beta} \cdot \sqrt\frac{V_R/\SI{}{m^3}}{\pi \: T_{60}/\SI{}{s}},
    \label{eq:dc}
\end{align}
where $\alpha$ and $\beta$ are the directivity factors of the acoustic source  and the microphone, respectively, and $V_R$ is the room volume. 
We consider omnidirectional microphones, i.e., $\beta = 1$.
The directivity factor of the source $\alpha$ is drawn from the uniform distribution $\mathcal{U}(2.5, 5.5)$, which corresponds to the interval around the directivity factor of the cardioid pattern, in order to account for fluctuations of the \gls{DRR} of recorded \glspl{RIR} for different positions.

Given the desired \gls{DRR} $\eta$, the  variance $\sigma_b^2$ of the Gaussian process $b(n)$ is chosen such that the \gls{DRR} of the \gls{RIR} $h(n)$, i.e.,
% \begin{align}
%     %\widehat{\eta} = \frac{\sum\limits_{n=N_d-w}^{N_d+w} h^2(n)}{\sum\limits_{n=N_d+w+1}^{N} h^2(n)},
%     \widehat{\eta} = \left(\sum\limits_{n=N_d-w}^{N_d+w} h^2(n) \right) \Bigg/ \left(\sum\limits_{n=N_d+w+1}^{N} h^2(n)\right),
%     \label{eq:drr}
% \end{align}
\begin{align}
    %\widehat{\eta} = \frac{\sum\limits_{n=N_d-w}^{N_d+w} h^2(n)}{\sum\limits_{n=N_d+w+1}^{N} h^2(n)},
    \widehat{\eta} = \frac{\sum\limits_{n=N_d-w}^{N_d+w} h^2(n) }{\sum\limits_{n=N_d+w+1}^{N} h^2(n)},
    \label{eq:drr}
\end{align}
matches the \gls{DRR} $\eta$, with $w$ defining the length of a small window around the impulse at delay $N_d$, which belongs to the direct path.
Here, we use $w {=} 40$ as proposed in \cite{Bryan20}.
% Since $h_I(n)$ also contributes to the power of the reverberation, i.e., the denominator in \eqref{eq:drr}, there is no closed form solution for calculating the  variance $\sigma_b^2$ of the Gaussian process $b(n)$.
% We solve this issue by iteratively adjusting the variance $\sigma_b^2$ until the \gls{DRR} $\widehat{\eta}$ of the \gls{RIR} $h(n)$  approximately meets the desired \gls{DRR} $\eta$.

%\begin{figure}[tb]
%	\centering
%	\resizebox{\linewidth}{!}{
%		\input{images/comp}
%    }
%   	\caption{Comparison of ISM, proposed method and example \gls{RIR} from MIRaGe %($T_{60} = \SI{0.6}{s}$, distance $\SI{1.8}{m}$).}	
    %\label{fig:comp}
%\end{figure}
%% Tex magic
% !TeX spellcheck = en_US
% !TeX encoding = utf-8
% !TeX root = ./../main.tex
% !TeX program = pdflatex 

\section{Distance Estimator} \label{Sec:DistEst}
We use our \gls{CRNN} from~\cite{distance_estimator} with the \gls{STFT} of the signals of a microphone pair as input for distance estimation. 
The \gls{STFT} is represented in the form of its absolute value and the sine and cosine of its phase for each microphone signal.
Note that the \gls{STFT} as input feature comes with the advantage that it does not only contain information about the source microphone distance in the form of the \gls{DRR}-related \glspl{ILD}, which can be derived from it, but also  useful side information for distance estimation, as discussed in \cite{distance_estimator}.
Before calculating the \gls{STFT}, all signals are normalized to the range $[-1, 1]$.
The model is trained to solve distance estimation as a classification problem with a class granularity of \SI{0.1}{m}.

Since only simulated data should be involved in the training procedure, also the best-performing checkpoint can only be determined based on an independent validation data set of simulated \glspl{RIR}.
However, the best-performing checkpoint for simulated data might not correspond to the best-performing checkpoint for real-world data.
We solve this issue via \gls{SWA}~\cite{izmailov2019averaging}.
Thereby, the model weights of the last \SI{25}{\%} of the checkpoints are averaged.
As mentioned in~\cite{cha2021swad}, this might also lead to flatter minima of the error plane, which can lead to a better generalization to other domains, e.g., a better generalization from simulated training data to real-world data.

\section{Experiments} \label{Sec:Experiments}
We simulated a data set of \SI{100}{k} \glspl{RIR} to train the distance estimator. 
Thereby, \SI{10}{k} different rooms are simulated. 
The length and width of the rooms are drawn from $\mathcal{U}\left(\SI{5}{m}, \SI{7}{m}\right)$ and their ceiling height from $\mathcal{U}\left(\SI{2.4}{m}, \SI{3.0}{m}\right)$.
Moreover, the sound decay times $T_{60}$ of the rooms were drawn from $\mathcal{U}\left(\SI{0.2}{s}, \SI{0.7}{s}\right)$

Ten constellations consisting of a source and a microphone pair with \SI{8}{cm} inter-microphone distance are generated for each room.
Therefore, the microphone pair was placed in the room with random position and orientation. 
Next, the source is placed relative to the microphone pair with a randomly drawn \gls{DoA} and distance so that a minimum distance of \SI{0.3}{m} and a maximum distance of \SI{5}{m} (or the largest possible distance that would fit into the area considered for source placement) was maintained.
Hereby, a minimal distance of \SI{0.5}{m} to the walls and \SI{1}{m} to the ceiling and floor is kept for each microphone and acoustic source. 
If the acoustic source would have to be placed outside the considered area for the drawn distance and \gls{DoA}, the \gls{DoA} is increased until the source position is within the considered area. 
The azimuth of the source's direction of view  is randomly drawn from $\mathcal{U}\left(-\SI{ 90}{\degree}, \SI{ 90}{\degree}\right)$ relative to the direct line of sight between the source and the microphone pair while the  corresponding elevation is randomly drawn from $\mathcal{U}\left(-\SI{15}{\degree}, \SI{15}{\degree}\right)$.
All simulated \glspl{RIR} have a length of $N{=}\SI{16384}{samples}$. 
The image source method was realized using \gls{PRA}~\cite{pra}.

In order to evaluate the ability of a distance estimator, which is trained solely using simulated \glspl{RIR}, to generalize to real-world data, we utilize two data sets of recorded \glspl{RIR}, namely MIRaGe \cite{MIRaGe} and MIRD \cite{mird} as test sets.
Both \gls{RIR} data sets were recorded in a room of size $\SI{6}{m} \times \SI{6}{m} \times \SI{2.4}{m}$ with configurable reverberation times. From the data sets, we selected only those microphone pairs that have a spacing of $\SI{8}{cm}$.
The acoustic sources of MIRD are placed on a regular grid at either \SI{1}{m} or \SI{2}{m} distance from a single microphone array with \glspl{DoA} between $\SI{\pm 90}{\degree}$ with $\SI{15}{\degree}$ steps in between.
Here, we used all examples with a sound decay time $T_{60}$ of \SI{360}{ms} and \SI{610}{ms}, which results in a total of $364$ test samples. 
In contrast, the MIRaGe data set has a cube-shaped volume, the so-called grid (46\SI{}{cm}$\times$36\SI{}{cm}$\times$32\SI{}{cm}), in which the sound source is positioned and from which we have selected $100$ positions. 
The microphone arrays are placed at defined distances (\SI{1}{m}, \SI{2}{m}, \SI{3}{m}), three facing the acoustic source and three at an angle of $\SI{45}{\degree}$. Additionally, $25$ outside of the grid (OOG) source positions are distributed across the room. From the available sound decay times $T_{60}$ we selected \SI{300}{ms} and \SI{600}{ms}, which resulted in $1200$ test samples for source positions from the grid and $300$ test samples for source positions outside of the grid.

Microphone signals with a length of \SI{1}{s} are generated by convolving clean speech from the LibriSpeech data set \cite{Librispeech} with the \glspl{RIR}.
During training the speech samples are randomly drawn from the train-clean-100 subset of LibriSpeech. 
For the evaluation, ten speech samples from the test-clean subset of LibriSpeech were used per constellation of source and microphone pair to mitigate the influence of the speech on the distance estimates.
Moreover, \gls{AWGN} is added to the microphone signals in order to simulate sensor noise with a \gls{SNR}, which is randomly drawn from $\mathcal{U}\left(\SI{40}{dB}, \SI{60}{dB}\right)$.

The distance estimators were trained for \SI{500}{k} iterations utilizing the Adam optimizer~\cite{Kingma14} with a batch size of $16$ and a learning rate of $3 {\cdot} 10 ^{-4}$.
Thereby, a checkpoint is created every \SI{10}{k} iterations.
The \gls{STFT} for feature extraction uses a Blackman window with a length of \SI{25}{ms} and shift of \SI{10}{ms}.

We evaluate the performance of the distance estimators by calculating the \gls{MAE} of the $I$ distance estimates per data set with 
\begin{align}
    \mathrm{MAE} = \frac{1}{I}\sum_{i=1}^I \big| d_i - \hat{d}_i\big|, 
\end{align}
where $d_i$ denotes the ground truth distance and $\hat{d}_i$ its estimate.
Note that estimated distance classes are mapped to the distance estimate $\hat{d}_i$ before calculating the \gls{MAE}.

\begin{table}
  \caption{Comparison of the proposed approach to \gls{RIR} simulation and the \acrfull{ISM} with different source directivity patterns. Additionally to the results on MIRD and MIRaGe results on a simulated version of MIRaGe (Sim.) are reported. We use the same approach to \gls{RIR} simulation for the simulated version of MIRaGe and the training of the corresponding distance estimator.} 
  \centering
  \begin{tabular}{c c c c c }
    \toprule
    \textbf{Method} & \textbf{Source Directivity} & \multicolumn{3}{c}{\textbf{MAE / \si{m}}} \\
    &  &\textbf{MIRD} &\textbf{MIRaGe}  &\textbf{Sim.}\\
    \midrule   
    ISM & Omnirectional & 0.75 &0.54 &0.20 \\
    ISM & Subcardioid & 0.51 &0.47 &0.17 \\
    ISM & Cardioid &0.27 &0.46 &0.16 \\
    ISM & Supercardiod &0.49&0.61 &0.21 \\
    ISM & Hypercardioid &0.32 &0.54 &0.21 \\
    \midrule   
    Proposed & Cardioid &0.20 &0.31 &0.26 \\
    \bottomrule
\end{tabular}
    \label{tab:pra_vs_hybrid}
\end{table}

\Cref{tab:pra_vs_hybrid} compares the performance of  a distance estimator trained with \glspl{RIR} which are simulated using the proposed method to the the performance of distance estimators trained with \glspl{RIR} which are simulated via the the image source method using different directivity patterns for the source.
It can be seen that the proposed \gls{RIR} simulation method leads to a significantly better distance estimation performance on MIRD and MIRaGe compared to the \acrfull{ISM}.
Moreover, it can be seen that the model which is trained with \glspl{RIR} from the image source method with cardioid source directivity exhibits the best performance of all models whose training data were generated using the image source method. 
In contrast, the distance estimators which are trained with \glspl{RIR} generated with less pronounced source directivities perform worst.

In addition to that, the distance estimators are evaluated on a simulated version of MIRaGe, which was generated by the same \gls{RIR} simulator as the one used to generate the training data for the respective model.
While there is a large gap between the performance on simulated and recorded \glspl{RIR} for distance estimators whose training data was simulated using the image source method, this gap is relatively small for a distance estimator trained with data for the proposed method.
This means that the proposed method improves the generalization ability of a distance estimator from simulated training data to real data by far. 

\begin{table}[b]
    \setlength{\tabcolsep}{1.mm}
    \caption{Comparison of the proposed approach to \gls{RIR} simulation and the \gls{DRR} augmentation technique from \cite{GeoJournal}, which scales the impulse belonging to the direct path with the factor $\alpha$. $\alpha$ is either randomly drawn as in \cite{GeoJournal} or calculated so that  the resulting \gls{RIR} meets the target \gls{DRR} from \eqref{hat_d}.}
    \centering
        \begin{tabular}{c c  c c c c c}
            \toprule
             \textbf{Method} & \textbf{Source Directivity} &\textbf{DRR aug.}  & \multicolumn{2}{c}{\textbf{MAE / \si{m}}} \\
            & & &\textbf{MIRD} &\textbf{MIRaGe} \\
            \midrule   
            ISM& Omnidirectional & $\alpha\sim \mathcal{U}\left(1, 3\right)$ & 0.27 & 0.56 \\
            ISM& Omnidirectional & $\alpha$ based on \eqref{hat_d}  & 0.23 & 0.43  \\
            ISM& Cardioid &  $\alpha$ based on \eqref{hat_d}  & 0.24 & 0.37  \\
            \midrule
            Proposed &Cardioid & - & 0.20 & 0.31 \\
            \bottomrule
        \end{tabular}
    \label{tab:ablation1}
\end{table}

Results for the distance estimation performance, which can be achieved by a combination of \glspl{RIR} from the image source method and the \gls{DRR} augmentation technique proposed in \cite{GeoJournal}, can be found in \cref{tab:ablation1}.
Thereby, the \gls{DRR} augmentation method varies the \gls{DRR} of the \glspl{RIR} by scaling the part of the \glspl{RIR} belonging to the direct path propagation.
Compared to the random scaling of the direct path component, which we proposed in \cite{GeoJournal} to increase the \gls{DRR}, better distance estimates can be achieved by scaling the direct path so that the \glspl{RIR} show a \gls{DRR} which is calculated based on \eqref{hat_d} as in the proposed method.
Hereby, the influence of the directivity on the direct path $D(\varphi, \varrho)$ in \eqref{hat_d} is calculated for the cardioid pattern. 
Further, the distance estimation performance is better when simulating sources with a cardioid directivity instead of omnidirectional sources. 
From this we hypothesize that the distance estimator benefits from incorporating the source's directivity into the model of the early specular reflections. 
However, the performance which can be achieved by using the proposed \gls{RIR} simulator cannot be reached.

\begin{table}[tb]
    \setlength{\tabcolsep}{1.2mm}
    \caption{Ablation study of the proposed approach to \gls{RIR} simulation by varying the source's directivity pattern, the maximum order of the image sources $K$ used to simulate $h_I(n)$ and the method used calculate the late part of the \glspl{RIR}. The last line corresponds to the proposed parametrization.}
    \centering
    \begin{tabular}{c c c c c c  c}
        \toprule
        \textbf{Source Directivity} &\textbf{Order $K$} & \textbf{Late RIR}  & \multicolumn{2}{c}{\textbf{MAE / \si{m}}} \\
        & & &\textbf{MIRD} &\textbf{MIRaGe} \\
        \midrule   
        Cardioid & 0 & Stochastic & 0.52 & 0.54 \\
        Omnidirectional & 3 & Stochastic  & 0.36 & 0.36 \\
        Cardioid & 3 & ISM & 0.24 & 0.42 \\
        \midrule
        Cardioid  & 3 & Stochastic  & 0.20 & 0.31 \\
        \bottomrule
    \end{tabular}
\label{tab:ablation2}
\end{table}
An ablation study for the proposed \gls{RIR} simulator is given in \cref{tab:ablation2}. 
It is shown that the distance estimation performance degrades a lot if only the direct path is simulated via the image source method, i.e., $K{=}0$, which again speaks for the importance of a correct simulation of the specular early reflections.
Moreover, it can be seen that the stochastic process from \eqref{eq:dc} models the diffuse reflections of higher order better than the image source method.
In addition to that,  simulating omnidirectional sources and also choosing $D(\varphi, \varrho){=}1$ in \eqref{hat_d} degrades the performance of the distance estimator.

\flushbottom
%% Tex magic
% !TeX spellcheck = en_US
% !TeX encoding = utf-8
% !TeX root = ./../main.tex
% !TeX program = pdflatex 

\section{Summary}\label{Sec:Conclusions}
In this paper, we presented a new approach to simulate \glspl{RIR} for the training of a DNN-based acoustic distance estimator to improve its performance in real-world scenarios.
Thereby, the image source method was utilized to simulate the reflections of lower order, which mainly show a specular character.
A cardioid pattern is used to simulate the source's directivity in the image source method because in real-world scenarios acoustic sources typically exhibit a directivity pattern, which largely differs from an omnidirectional directivity pattern.
In addition to that, the mainly diffuse reflections of higher order are modeled via an exponentially decaying stochastic process.
The power of the latter is scaled such that the \gls{DRR} of the \gls{RIR} fits to the distance between the source and the microphone. 
Experiments on recorded \glspl{RIR} show that our contribution improves the simulated training data of a distance estimator to match the characteristics present in real data better than previous approaches.

In future works we will investigate the suitability of the proposed approach to \gls{RIR} simulation to generate training data for data-driven models for other purposes, like dereverberation, speech enhancement or room parameter estimation, e.g., \gls{DRR} and sound decay time $T_{60}$ estimation.

% In this paper, we presented a new approach how to simulate RIRs for the training of a distance estimator to improve its performance in real-world scenarios.
% The training is performed solely on simulated data and the performance is evaluated on real-world data, whereby our contribution improves the training data  in a way that the characteristics present in real data are better matched as by previous approaches. Therefore, the test-train mismatch is eased up to a certain degree, and distance estimation for downstream tasks gets more robust.
% Two open challenges remain: Firstly, an evaluation of the performance on a data set with a much more diverse collection of environments to check if the model has generalization capability towards other rooms with other reverberation times. Secondly, overfitting remains a persistent, albeit mitigated, phenomenon as the results indicate that the performance might be boosted if the model selection may be based on different criteria.
% \fromjs{In future work we will study if the proposed approach is beneficial to speech enhancement in reverberant environments. Furthermore, the software will be published as open source software on our repository \cite{GitHubPaderWASN}.}

%\section*{Acknowledgment}
%Funded by the Deutsche Forschungsgemeinschaft (DFG, German Research Foundation) - Project 282835863.

\pagebreak
\bibliographystyle{IEEEbib}
\bibliography{main}

\end{document}